\documentclass[amsmath]{aastex702}
\usepackage{bm}
\usepackage{graphicx}

\shorttitle{Revisiting damping of Alfv{\'e}n waves}
\shortauthors{Yutaka Ohira}
\usepackage{natbib}
\usepackage{lineno}
\linenumbers
\begin{document}

\title{Revisiting damping of Alfv{\'e}n waves in partially ionized plasmas}
\author[0000-0002-2387-0151]{Yutaka Ohira}
\affiliation{Department of Earth and Planetary Science, The University of Tokyo, \\
7-3-1 Hongo, Bunkyo-ku, Tokyo 113-0033, Japan}
\email{y.ohira@eps.s.u-tokyo.ac.jp}

\begin{abstract}
We investigate the damping rate of Alfv{\'e}n waves by means of the linear analysis of the two-fluid ten-moments system. 
To treat effects of viscosity in a wide frequency range, we take account of the time evolution of the traceless part of the pressure tensor. 
In addition, to treat effects of viscosity in a wide range of the ionization degree, we consider interactions between hydrogen atoms and between protons and hydrogen atoms in the equation of the traceless part of the pressure tensor. 
We find approximate analytical solutions to the dispersion relation for some limits.   
Our analysis in the low-frequency limit shows that the viscous damping dominates over the frictional damping in the high-beta plasma, which is independent of the ionization degree.
Therefore, the viscosity of hydrogen gas significantly affects cosmic-ray propagation in a partially ionized high-beta plasma. 
\end{abstract}

\keywords{\uat{Cosmic rays}{329} --- \uat{Plasma physics}{2089} --- \uat{Interstellar plasma}{851} --- \uat{Plasma astrophysics}{1261} --- \uat{Ion-neutral reactions}{2263} --- \uat{Magnetohydrodynamics}{1964}}

\section{Introduction}
\label{sec1}
Since motion of charged cosmic rays (CRs) depends on the strength of the magnetic field and its fluctuations, 
to understand the CR acceleration and propagation, we need to know how the magnetic field is fluctuated in a system. 
In a uniform magnetic field, charged particles are confined to the magnetic field line by gyro motion, but they can move freely along the magnetic field line. 
On the other hand, in a disordered magnetic field, particle trajectories become complex and, over sufficiently long timescales, can be regarded as random motion.
The random motion enables particles to traverse magnetized plasmas moving with different velocities, so that particles are accelerated in the shock \citep{axford77,krymsky77,bell78,blandford78}, shear flow \citep{berezhko81,earl88,webb89,ostrowski90,rieger06}, and turbulence \citep[e.g][]{skilling75,schlickeiser98,ohira13a,lemoine21}. 
Furthermore, the diffusion of CRs controls when they escape from the acceleration site \citep{ptuskin05,ohira10}, and how long CRs stay around the acceleration site \citep{aharonian96,gabici09,fujita09,li10,ohira11,ohira25} and in our galaxy \citep{kulsrud69}.
Because the efficiencies of acceleration and diffusion of CRs are strongly depends on the magnetic field fluctuation, 
the generation and damping of the magnetic field fluctuation in various systems have been intensively studied for long time. 

Most gas is in a partially ionized state in the equatorial plane of a galaxy. 
In addition, gamma rays and line emissions of ionized molecules are observed from molecular clouds, where CRs produce the gamma rays and ionize molecules. 
To extract correct information from the observed gamma rays and line emissions of ionized molecules, 
we have to understand the CR propagation in the partially ionized medium accurately. 
The magnetic field fluctuation is driven by some astrophysical activities (e.g. stellar wind and supernova explosion) and the CR streaming instability. 
It is thought that a quasi-steady state of magnetic turbulence is achieved by balancing the generation mechanism with some cascade or damping processes. 

In the cosmic-ray community, it has long been believed  that the damping of magnetohydrodynamic (MHD) waves in partially ionized plasma is primarily caused by friction between neutral particles and ions \citep{piddington56,kulsrud69,brahimi20,recchia22}. 
\citet{kulsrud69} evaluated the viscosity of ions using the Branginskii model, concluding that it is negligible. 
The wave damping in the partially ionized chromosphere is also an active topic of research in solar physics \citep{soler24}. 
Regarding the role of viscosity, \citet{soler15} showed that its effect on chromospheric Alfv{\'e}n waves is not relevant.
On the other hand, in studies of interstellar MHD turbulence, in addition to friction, viscosity due to hydrogen-hydrogen scattering is also considered as the damping process of MHD waves \citep{lazarian04,xu15,xu16}. 
In addition to the hydrogen-hydrogen collision, the proton-hydrogen collision also contributes to the viscosity of hydrogen gas, but it was not considered in the above studies. 
Furthermore, the viscosity term is phenomenologically introduced only in the equation of motion of hydrogen gas. 
However, as the wave frequency approaches the scattering frequency from the low-frequency side, the analytical expression for the viscous term in the equation of motion becomes invalid. 
Since the cross section for viscosity is usually smaller than the momentum transfer cross section concerning friction \citep{vranjes13}, the validity of the analytical expression for the viscous term is lost before the frictional interaction becomes ineffective.
Moreover, when the scattering frequency is larger than the proton cyclotron frequency, the viscosity of protons could be comparable to that of hydrogen. In this case, the viscosity of protons must also be taken into account.

To address the above issues, it is necessary to derive effects of viscosity based on the kinematic framework, rather than simply adding a viscous term to the equations of motion in a phenomenological manner. 
Studies on viscosity based on kinematics began with Chapman and Enskog \citep{chapman53}, and continue to this day \citep[e.g][]{hunana22}. 
The effects of viscosity have been derived using various methods based on the kinematic theory.
In this work, we investigate the viscous and frictional damping rates of MHD waves by performing a linear analysis of the two-fluid (protons and hydrogen atoms), ten-moment-equation system derived by the Grad method \citep{grad49a,grad49b,grad58}. 
This study focuses specifically on Alfv{\'e}n waves that propagate parallel to a magnetic field line. 
In order to perform a linear analysis of general MHD modes, it is necessary to deal with the thirteen moment equation system that incorporates heat conduction. 
However, due to the complexity of the issue, it will be reserved for a future study.

We first introduce the two-fluid, ten-moment-equation system in Section \ref{sec2}. 
In Section \ref{sec3}, a linear analysis is performed in order to derive the dispersion relation for Alfv{\'e}n waves propagating along a magnetic field line in a homogeneous system. 
In section \ref{sec4}, we give the decay rate by numerically solving the dispersion relation. 
Analytical approximate solutions are derived for some limiting cases. 
Finally, we discuss our results and summarize this work in Section \ref{sec5}.

\section{Two-fluid, ten-monent-equation system}
\label{sec2}
The viscous term of fluid equations was derived by Chapman and Enskog on the basis of the kinematic theory \citep{chapman53}. 
Subsequently, \citet{braginskii58,braginskii65} applied the Chanpman-Enskog method to magnetized plasma systems and showed that viscosity and thermal conductivity depend on the direction relative to the magnetic field line. 
Several methodologies for deriving the viscous term have been proposed since then \citep{hunana22,takamoto10}. 
In this work, we use the Grad method \citep{grad49a,grad49b,grad58} to derive the effect of viscosity on the multi-fluid system. 
The starting point is the Boltzmann equation, 
\begin{equation}
\frac{\partial f_s }{\partial t} + {\bm v} \cdot {\bm \nabla} f_s +\frac{q_s}{m_s} \left( {\bm E} + \frac{{\bm v}}{c}\times{\bm B} \right) \cdot {\bm \nabla}_v f_s= \left(\frac{\delta f_s }{\delta t}\right)_{\rm c}
\label{eq:b}
\end{equation}
where $f_s, v, q_s, m_s, c, {\bm E}, {\bm B}~{\rm and}~ (\delta f_s /\delta t)_{\rm c}$ are the distribution function, velocity, charge, mass, speed of light, electric field, magnetic field, and binary collision term, respectively. 
The subscript, $s$, represents particle species, where $s={\rm e, p, and~H}$ for electron, proton, and hydrogen atom, respectively. 
If the distribution function is a shifted Maxwell distribution, the Euler equation without viscosity and thermal conductivity is derived by taking moments of the Boltzmann equation.
Thus, the effects of viscosity and thermal conductivity originate from the deviation from the shifted Maxwell distribution. 
In the Chapman-Enskog method, the deviation is obtained by solving the collisional Boltzmann equation in the iterative method. 
In the Grad method, the distribution function is expanded in tensorial Hermite polynomial, where the coefficients of the Hermite polynomial expansion are given by moments of the distribution function. 
In this work, we consider only incompressible Alfv{\'e}n waves in a uniform system as a first step. Therefore, the thermal conduction can be ignored because temperature is uniform.  
In this case, the Hermite polynomial expansion of the distribution function is given by 
\begin{equation}
f_s = n_s\left(\frac{m_s}{2\pi k_{\rm B}T_s}\right)^{3/2}\exp\left(-\frac{m_s w_s^2}{2k_{\rm B}T_s}\right) \left(1+\frac{m_s}{2k_{\rm B}T_s p_s} {\bm \tau}_s:{\bm w}_s{\bm w}_s\right) ;~ ~ ~ ~ {\bm w}_s={\bm v}-{\bm u}_s,
\label{eq:hermite}
\end{equation}
where $k_{\rm B}$ is the Boltzmann constant, and ${\bm w}_s$ is the particle velocity in the rest frame of the $s$ fluid. 
$n_s, {\bm u}_s, T_s=p_s/n_s k_{\rm B}, p_s$, and ${\bm \tau}_s$ are the number density, mean velocity, temperature, trace part of the pressure tensor, and traceless part of the pressure tensor. 
These are given by the following velocity moments of the distribution function. 
\begin{equation}
n_s = \int f_s {\rm d}^3 v;~ ~ ~ ~ ~{\bm u}_s = \frac{1}{n_s}\int {\bm v}f_s {\rm d}^3 v;~ ~ ~ ~ ~p_s = \frac{m_s}{3}\int w_s^2f_s {\rm d}^3 v;~ ~ ~ ~ ~{\bm \tau}_s = m_s \int \left({\bm w}_s {\bm w}_s - \frac{w_s^2}{3}{\bf I}\right)f_s {\rm d}^3 v, 
\end{equation}
%
%
where ${\bf I}$ is the unit tensor. Since ${\bm \tau_s}$ is a symmetric tensor, there are ten independent moments for each species, $s$. 
Taking the velocity moments of the Boltzmann equation, Eq. (\ref{eq:b}), we obtain the following equations for the ten moments: 
\begin{equation}
\frac{D_s n_s}{D t}+n_s {\bm \nabla} \cdot {\bm u}_s = 0,
\label{eq:ns}
\end{equation}
\begin{equation}
m_sn_s\frac{D_s {\bm u}_s}{D t}+{\bm \nabla} p_s + {\bm \nabla} \cdot {\bm \tau}_s -q_sn_s\left({\bm E}+\frac{{\bm u}_s}{c}\times {\bm B} \right)= \dot{{\bm M}}_{{\rm c},s},
\label{eq:us}
\end{equation}
\begin{equation}
\frac{3}{2}\frac{D_s p_s}{D t}+\frac{5}{2}p_s({\bm \nabla} \cdot {\bm u}_s) + {\bm \tau}_s: {\bm \nabla} {\bm u}_s= \dot{E}_{{\rm c},s}, 
\label{eq:ps}
\end{equation}
\begin{eqnarray}
\frac{D_s {\bm \tau}_s}{D t}+{\bm \tau}_s {\bm \nabla} \cdot {\bm u}_{\rm s} +({\bm \Omega}_s\times {\bm \tau}_s-{\bm \tau}_s\times {\bm \Omega}_s) &+& p_s \left\{ {\bm \nabla} {\bm u}_s + ({\bm \nabla} {\bm u}_s)^{\rm T}-\frac{2}{3}({\bm \nabla}\cdot {\bm u}_s){\bf I} \right\} \nonumber \\
&+& {\bm \tau}_s\cdot{\bm \nabla} {\bm u}_s + ({\bm \tau}_s\cdot{\bm \nabla} {\bm u}_s)^{\rm T}-\frac{2}{3}({\bm \tau}_s:{\bm \nabla} {\bm u}_s){\bf I} = \dot{\bm \tau}_{{\rm c},s}, 
\label{eq:taus}
\end{eqnarray}
where $D_s/Dt=\partial/\partial t + {\bm u}_s\cdot {\bm \nabla}$, ${\bm \Omega_{s}}=q_s{\bm B}/m_sc$. 
The terms on the right hand side, $\dot{{\bm M}}_{{\rm c},s}, \dot{E}_{{\rm c},s}$, and $\dot{\bm \tau}_{{\rm c},s}$, are velocity moments of collision term. 
For the Coulomb and hard sphere interactions with a low relative bulk velocity limit ($|{\bm u}_s-{\bm u}_t|\ll (3k_{\rm B}T_s/m_s)^{1/2}, (3k_{\rm B}T_t/m_t)^{1/2}$), they are given by
\begin{eqnarray}
\dot{\bm M}_{{\rm c},s}&=&\int m_s{\bm v}\left(\frac{\delta f_s }{\delta t}\right)_{\rm c} {\rm d}^3v=-\sum_{t}m_{\rm s}n_{\rm s}\nu_{\rm st}({\bm u}_{\rm s}-{\bm u}_{\rm t}) \\
\dot{E}_{{\rm c},s}&=&\int \frac{1}{2}m_s w_s^2\left(\frac{\delta f_s }{\delta t}\right)_{\rm c} {\rm d}^3v=-\sum_t \frac{3m_s n_s\nu_{st}}{m_s+m_t}k_{\rm B}\left(T_s-T_t\right) \\
\dot{\bm \tau}_{{\rm c},s} &=& \int m_s\left( {\bm w}_s{\bm w}_s - \frac{w_s^2}{3}{\bf I}  \right)\left(\frac{\delta f_s }{\delta t}\right)_{\rm c} {\rm d}^3v=-\sum_{t \neq s} \frac{2m_s \nu_{st}}{m_s+m_t}\left\{ {\bm \tau}_s -\frac{n_s}{n_t} {\bm \tau}_t + \frac{3}{5}\frac{m_t}{m_s}\left({\bm \tau}_s + \frac{n_s m_s}{n_t m_t}{\bm \tau}_t \right)\right\} - \frac{6}{5}\nu_{ss}{\bm \tau}_s, 
\label{eq:tauc}
\end{eqnarray}
where $\nu_{st}$ is the collision frequency between particles $s$ and $t$ \cite[for details, see][]{burgers69, schunk75}.
The time evolution of electromagnetic fields, ${\bm E}$ and ${\bm B}$, is given by the Maxwell equations. 
Then, we obtain the closed equation system. 

Since we focus on Alfv{\'e}n waves in this work, the frequency range can be limited to a lower frequency than the plasma frequency.  
In addition, the electrical resistivity does not play an important role in the damping of the Alfv{\'e}n waves as long as the wave length is sufficiently longer than the electron inertial length. 
Therefore, the electron inertia, displace current, and resistivity can be ignored in this system, that is, the ideal MHD approximation can be applied. 
Then, Maxwell's equations are reduced to 
\begin{equation}
{\bm E} = -\frac{{\bm u}}{c}\times {\bm B};~ ~ ~ ~\frac{\partial {\bm B}}{\partial t} ={\bm \nabla} \times ({\bm u}\times {\bm B});~ ~ ~ ~{\bm \nabla} \times {\bm B}=\frac{4\pi}{c}(q_{\rm p}n_{\rm p}{\bm u}_{\rm p} + q_{\rm e}n_{\rm e}{\bm u}_{\rm e}).
\label{eq:mhd}
\end{equation}
To treat protons and electrons as a single fluid, we define the following quantities; 
\begin{equation}
\rho = m_{\rm p}n_{\rm p} + m_{\rm e}n_{\rm e};~ ~ ~ ~{\bm u}=(\rho_{\rm p}{\bm u}_{\rm p}+\rho_{\rm e}{\bm u}_{\rm e})/\rho;~ ~ ~ ~p=p_{\rm p}+p_{\rm e};~ ~ ~ ~{\bm \tau} = {\bm \tau}_{\rm p} + {\bm \tau}_{\rm e}.
\end{equation}
From Eqs.(\ref{eq:ns})-(\ref{eq:ps}) for protons and electrons, fluid equations for the plasma component are  
\begin{equation}
\frac{D \rho}{D t}+\rho {\bm \nabla} \cdot {\bm u} = 0,  
\label{eq:rho}
\end{equation}
\begin{equation}
\rho\frac{D {\bm u}}{D t}+{\bm \nabla} p + {\bm \nabla} \cdot {\bm \tau} -\frac{({\bm \nabla}\times{\bm B}) \times {\bm B}}{4\pi} = -\rho\nu_{\rm pH}({\bm u}-{\bm u}_{\rm H}),
\label{eq:u}
\end{equation}
\begin{equation}
\frac{3}{2}\frac{D p}{D t}+\frac{5}{2}p({\bm \nabla} \cdot {\bm u}) + {\bm \tau}: {\bm \nabla} {\bm u}= -3\rho \left(\frac{1}{2}\frac{p}{\rho} -\frac{p_{\rm H}}{\rho_{\rm H}} \right) \left( \frac{1}{2}\nu_{\rm pH}+\frac{m_{\rm e}}{m_{\rm p}}\nu_{\rm eH}\right),
\label{eq:p}
\end{equation}
where $m_{\rm p}\gg m_{\rm e}, n_{\rm p}\approx n_{\rm e}, u\approx u_{\rm p}$, and $p\approx 2p_{\rm p}\approx 2p_{\rm e}$ are assumed, as is commonly used in the standard MHD approximation. 
In addition, $m_{\rm p}\approx m_{\rm H}$ is also assumed. 

The equations for the traceless part of the pressure tensor, Eqs.(\ref{eq:taus}) and (\ref{eq:tauc}), show that ${\bm \tau}_{\rm p}$ and ${\bm \tau}_{\rm e}$ evolve in the different time scale due to the terms with ${\bm \Omega}_s$ and $\dot{\bm \tau}_{{\rm c},s} $. 
Consequently, it is not possible to express the equations of ${\bm \tau}_{\rm p}$ and ${\bm \tau}_{\rm e}$ as a single equation of ${\bm \tau}={\bm \tau}_{\rm p}+{\bm \tau}_{\rm e}$. 
Therefore, in a general case, the equations of ${\bm \tau}_{\rm p}$ and ${\bm \tau}_{\rm e}$ have to be solved separately. 
However, we can approximately ignore effects of ${\bm \tau}_{\rm e}$ on the single plasma fluid (electrons + protons) as long as the time scale is longer than the electron gyro period ($t>\Omega_{\rm e}^{-1}$) and the electron temperature is not so high, $T_{\rm e}\lesssim T_{\rm p}$. 
The cyclotron frequency and scattering frequencies of electrons are much larger than those of protons ($\Omega_{\rm e}\gg\Omega_{\rm p}, \nu_{\rm ee}\gg\nu_{\rm pp}, \nu_{\rm ep}\gg\nu_{\rm pe}$, and $\nu_{\rm eH}\gg\nu_{\rm pH}$) for $T_{\rm e}\lesssim T_{\rm p}$. 
In this case with $t>\Omega_{\rm p}^{-1}$, the traceless part of the pressure tensor is roughly estimated by $\tau_s \approx p_s u/L\Omega_s$ or $p_s u/L\nu_{st}$, so that we can assume $\tau \approx \tau_{\rm p}$, where $L$ is the gradient length scale of the velocity field, $u$. 
Then, the equations for $\tau$ is given by one for $\tau_{\rm p}$ 
\begin{eqnarray}
\frac{D {\bm \tau}}{D t}&+&{\bm \tau} {\bm \nabla} \cdot {\bm u} +({\bm \Omega}_{\rm p}\times {\bm \tau}-{\bm \tau}\times {\bm \Omega}_{\rm p}) + p_{\rm p} \left\{ {\bm \nabla} {\bm u} + ({\bm \nabla} {\bm u})^{\rm T}-\frac{2}{3}({\bm \nabla}\cdot {\bm u}){\bf I} \right\} \nonumber \\
&+& {\bm \tau}\cdot{\bm \nabla} {\bm u} + ({\bm \tau}\cdot{\bm \nabla} {\bm u})^{\rm T}-\frac{2}{3}({\bm \tau}:{\bm \nabla} {\bm u}){\bf I} = - \left(\frac{6}{5}\nu_{\rm pp} +\frac{8}{5}\nu_{\rm pH} \right) {\bm \tau} + \frac{2}{5}\nu_{\rm Hp}{\bm \tau}_{\rm H}, 
\label{eq:tau}
\end{eqnarray}
where $m_{\rm p}\gg m_{\rm e}$ is again used. 
Fluid equations for hydrogen atoms are given by 
\begin{equation}
\frac{D_{\rm H} \rho_{\rm H}}{D t}+\rho_{\rm H} {\bm \nabla} \cdot {\bm u}_{\rm H} = 0
\end{equation}
\begin{equation}
\rho_{\rm H}\frac{D_{\rm H} {\bm u}_{\rm H}}{D t}+{\bm \nabla} p_{\rm H} + {\bm \nabla} \cdot {\bm \tau}_{\rm H}= \rho\nu_{\rm pH}({\bm u}-{\bm u}_{\rm H}),
\end{equation}
\begin{equation}
\frac{3}{2}\frac{D_{\rm H} p_{\rm H}}{D t}+\frac{5}{2}p_{\rm H}({\bm \nabla} \cdot {\bm u}_{\rm H}) + {\bm \tau}_{\rm H}: {\bm \nabla} {\bm u}_{\rm H}= 3\rho \left(\frac{1}{2}\frac{p}{\rho} -\frac{p_{\rm H}}{\rho_{\rm H}} \right) \left( \frac{1}{2}\nu_{\rm pH}+\frac{m_{\rm e}}{m_{\rm p}}\nu_{\rm eH}\right),
\label{eq:pH}
\end{equation}
\begin{eqnarray}
\frac{D_{\rm H} {\bm \tau}_{\rm H}}{D t}&+&{\bm \tau}_{\rm H} {\bm \nabla} \cdot {\bm u}_{\rm H} + p_{\rm H}\left\{ {\bm \nabla} {\bm u}_{\rm H} + ({\bm \nabla} {\bm u}_{\rm H})^{\rm T}-\frac{2}{3}({\bm \nabla}\cdot {\bm u}_{\rm H}){\bf I} \right\} \nonumber \\
&+& {\bm \tau}_{\rm H}\cdot{\bm \nabla} {\bm u}_{\rm H} + ({\bm \tau}_{\rm H}\cdot{\bm \nabla} {\bm u}_{\rm H})^{\rm T}-\frac{2}{3}({\bm \tau}_{\rm H}:{\bm \nabla} {\bm u}_{\rm H}){\bf I} =  \frac{2}{5}\nu_{\rm pH}{\bm \tau}- \left(\frac{6}{5}\nu_{\rm HH} +\frac{8}{5}\nu_{\rm Hp} \right) {\bm \tau}_{\rm H}, 
\label{eq:tauH}
\end{eqnarray}
where $m_{\rm H}\approx m_{\rm p}\gg m_{\rm e}$ is again assumed. 
The Eqs. (\ref{eq:mhd}) - (\ref{eq:tauH}) are basic equations in this work.

If we are interested in physics with a time scale sufficiently longer than all the scattering time scale, $\nu_{\rm st}^{-1}$, we can approximately omit the time derivative of the traceless part of the pressure tensor, $D\tau/Dt$ and $D_{\rm H}\tau_{\rm H}/Dt$. 
Then, by using Eqs.(\ref{eq:tau}) and (\ref{eq:tauH}), we can eliminate the traceless part of the pressure tensor from this equation system, so that the Navier-Stokes equations are obtained. 
${\bm \nabla} \cdot {\bm \tau}$ and ${\bm \nabla} \cdot {\bm \tau}_{\rm H}$ correspond to the viscous terms. 
Since we are interested in a wide frequency range in this work, we take account of the time derivative of the traceless part of the pressure tensor. 
\section{Linear analysis}
\label{sec3}
In this section, we perform linear analysis to derive the dispersion relation for Alfv{\'e}n waves in a uniform system, ${\bm u}={\bm u}_{\rm H}={\bm 0}, T=T_{\rm H}=T_0, \rho=\rho_{\rm 0}, \rho_{\rm H}=\rho_{\rm H,0},{\bm \tau}={\bm \tau}_{\rm H} = {\bf 0}$, and ${\bm B}=B_0 {\bm e}_x$. 
In this work, we focus on the Alfv{\'e}n wave propagating along the uniform background magnetic field. Therefore, the perturbation mode is incompressible ($\delta \rho=\delta \rho_{\rm H}=\delta p=\delta p_{\rm H}=0$), but perturbed quantities of $\delta {\bm u}, \delta {\bm u}_{\rm H}, \delta {\bm \tau}, \delta {\bm \tau}_{\rm H}$, and $\delta {\bm B}$ are nonzero and assumed to be proportional to $\exp \{i(\omega t - k x)\}$.
In this assumption, a mode with $\rm{Im}[\omega]>0$ is a damping mode and the damping rate is $\gamma_{\rm d}=\rm{Im}[\omega]$. 
Then, the linearized equations are given by
\begin{equation}
i\omega \rho_0\delta{\bm u} -i {\bm k} \cdot \delta {\bm \tau} - \frac{ik^2B_0^2}{4\pi \omega} \delta{\bm u} = -\rho_0\nu_{\rm pH}(\delta {\bm u}-\delta {\bm u}_{\rm H})
\label{eq:lup}
\end{equation}
\begin{equation}
i\omega \rho_{\rm H,0}\delta{\bm u}_{\rm H} -i {\bm k} \cdot \delta {\bm \tau}_{\rm H} = \rho_0\nu_{\rm pH}(\delta {\bm u}-\delta {\bm u}_{\rm H})
\end{equation}
\begin{equation}
i\omega {\bm k} \cdot \delta {\bm \tau} - ({\bm k} \cdot \delta {\bm \tau})\times {\bm \Omega}_{\rm p} - ik^2p_{\rm p,0} \delta {\bm u} = -\nu_{\rm v,pp}{\bm k} \cdot \delta {\bm \tau}+\nu_{\rm v,pH}{\bm k} \cdot \delta {\bm \tau}_{\rm H}
\label{eq:ltaup}
\end{equation}
\begin{equation}
i\omega {\bm k} \cdot \delta {\bm \tau}_{\rm H} - ik^2p_{\rm H,0} \delta {\bm u}_{\rm H} = \nu_{\rm v,Hp}{\bm k} \cdot \delta {\bm \tau}-\nu_{\rm v,HH}{\bm k} \cdot \delta {\bm \tau}_{\rm H}
\end{equation}
where $\nu_{\rm v,pp}, \nu_{\rm v,pH}, \nu_{\rm v,HH}$, and $\nu_{\rm v,Hp}$ are 
\begin{equation}
\nu_{\rm v,pp}=\frac{6}{5}\nu_{\rm pp} + \frac{8}{5}\nu_{\rm pH};~ ~ ~ ~\nu_{\rm v,pH}= \frac{2}{5} \left(\frac{\rho_0}{\rho_{\rm H,0}}\right)\nu_{\rm pH};~ ~ ~ ~\nu_{\rm v,HH}= \frac{6}{5}\nu_{\rm HH} +\frac{8}{5}\nu_{\rm Hp};~ ~ ~ ~\nu_{\rm v,Hp}= \frac{2}{5} \left(\frac{\rho_{\rm H,0}}{\rho_0}\right)\nu_{\rm Hp},
\end{equation}
and $\Omega_{\rm p}$ and $\nu_{st}$ are given by the background quantities. 
Then, the dispersion relation is given by the following sixth-degree polynomial in $\omega$, 
\begin{equation}
D=\sum_{i=0}^6 a_{i}\omega^{i}=0, 
\label{eq:fd}
\end{equation}
where the coefficients, $a_i$, are as follows.
\begin{eqnarray}
a_0&=& - \left[ \left( \nu_{\rm v,pp}^2 + \Omega_{\rm p}^2 \right)c_{\rm s}^2k^2  + \left( \frac{\rho_0}{\rho_{\rm H,0}} \right) \left\{ \nu_{\rm v,HH}\left(\nu_{\rm v,pp}^2+\Omega_{\rm p}^2 \right) -\nu_{\rm v,pp}\nu_{\rm v,pH}\nu_{\rm v,Hp}  \right \}\nu_{\rm pH} \right] v_{\rm A}^2k^2,  \\
a_1&=& -i\left[ \left\{ \left( \nu_{\rm v,pp}^2+\Omega_{\rm p}^2 +\nu_{\rm v,pp}\nu_{\rm v,pH}\right)\nu_{\rm pH}+\nu_{\rm v,pp}c_{\rm s}^2k^2 \right\}c_{\rm s}^2k^2 + \left\{ \nu_{\rm v,HH}\left(\nu_{\rm v,pp}^2+\Omega_{\rm p}^2\right)-\nu_{\rm v,pp}\nu_{\rm v,pH}\nu_{\rm v,Hp} +2\nu_{\rm v,pp}c_{\rm s}^2k^2 \right \}v_{\rm A}^2k^2 \right. \nonumber \\
&&\ \ \ \ \ \left.+ \left( \frac{\rho_0}{\rho_{\rm H,0}} \right) \left\{ \nu_{\rm v,pp}(\nu_{\rm v,HH}+\nu_{\rm v,Hp})c_{\rm s}^2k^2 +\left(\nu_{\rm v,pp}^2+\Omega_{\rm p}^2+2\nu_{\rm v,pp}\nu_{\rm v,HH}-\nu_{\rm v,pH}\nu_{\rm v,Hp}\right)v_{\rm A}^2k^2 \right \} \nu_{\rm pH} \right],\\
a_2&=& ( \nu_{\rm v,pp}^2+\Omega_{\rm p}^2+\nu_{\rm v,pp}\nu_{\rm v,HH} +c_{\rm s}^2k^2 )c_{\rm s}^2k^2 + ( \nu_{\rm v,pp}^2+\Omega_{\rm p}^2+2\nu_{\rm v,pp}\nu_{\rm v,HH} -\nu_{\rm v,pH}\nu_{\rm v,Hp} +c_{\rm s}^2k^2   )v_{\rm A}^2k^2 \nonumber \\
&&+\left\{ \nu_{\rm v,pp}^2\nu_{\rm v,HH}-\nu_{\rm v,pp}\nu_{\rm v,pH}\nu_{\rm v,Hp}  + \nu_{\rm v,HH}\Omega_{\rm p}^2 +(\nu_{\rm v,pH}+2\nu_{\rm v,pp})c_{\rm s}^2k^2 \right \}\nu_{\rm pH} \nonumber \\
&&+ \left(\frac{\rho_0}{\rho_{\rm H,0}}\right)\left\{ \nu_{\rm v,HH}(\nu_{\rm v,pp}^2+\Omega_{\rm p}^2) -\nu_{\rm v,pp}\nu_{\rm v,pH}\nu_{\rm v,Hp}  + ( \nu_{\rm v,pp}+\nu_{\rm v,HH}+\nu_{\rm v,Hp})c_{\rm s}^2k^2 + ( 2\nu_{\rm v,pp}+\nu_{\rm v,HH} )v_{\rm A}^2k^2 \right\}\nu_{\rm pH},\\
a_3&=& -i\left[ \nu_{\rm v,pp}\nu_{\rm v,pH}\nu_{\rm v,Hp} -(3\nu_{\rm v,pp}+\nu_{\rm v,HH})c_{\rm s}^2k^2 -(2\nu_{\rm v,pp}+\nu_{\rm v,HH})v_{\rm A}^2k^2 -\nu_{\rm v,HH}(\nu_{\rm v,pp}^2+\Omega_{\rm p}^2)  \right. \nonumber \\
&&\ \ \ \ \ + (\nu_{\rm v,pH}\nu_{\rm v,Hp}-2\nu_{\rm v,pp}\nu_{\rm v,HH}-\nu_{\rm v,pp}^2-\Omega_{\rm p}^2-c_{\rm s}^2k^2)\nu_{\rm pH}  \nonumber \\
&&\ \ \ \ \ \left. +\left(\frac{\rho_0}{\rho_{\rm H,0}}\right)\left\{\nu_{\rm v,pH}\nu_{\rm v,Hp}-2\nu_{\rm v,pp}\nu_{\rm v,HH}-\nu_{\rm v,pp}^2-\Omega_{\rm p}^2 -(c_{\rm s}^2+v_{\rm A}^2)k^2 \right\}\nu_{\rm pH} \right], \\
a_4&=& -(2c_{\rm s}^2+v_{\rm A}^2)k^2-(2\nu_{\rm v,pp}+\nu_{\rm v,HH})\nu_{\rm pH} - \nu_{\rm v,pp}^2 - 2\nu_{\rm v,pp}\nu_{\rm v,HH} + \nu_{\rm v,pH}\nu_{\rm v,Hp}-\Omega_{\rm p}^2 +\left(\frac{\rho_0}{\rho_{\rm H,0}}\right)(2\nu_{\rm v,pp}+\nu_{\rm v,HH})\nu_{\rm pH},\\
a_5&=& -i\left\{ \nu_{\rm pH}+2\nu_{\rm v,pp}+\nu_{\rm v,HH}+\left(\frac{\rho_0}{\rho_{\rm H,0}}\right)\nu_{\rm pH}\right\}, \\
a_6&=& 1,
\end{eqnarray}
where $c_{\rm s}=\sqrt{p_0/\rho_0}=\sqrt{p_{\rm H,0}/\rho_{\rm H,0}}$ and $v_{\rm A}=B_0/\sqrt{4\pi \rho_0}$ are 
the sound and Alfv{\'e}n velocites, respectively. 
The numerical solutions to Eq.(\ref{eq:fd}) are given in Figs. \ref{fig1}-\ref{fig3} for some cases. 
The approximate analytical solution in some limits are given in the next section. 

\section{Damping rate of Alfv{\'e}n waves in the partially ionized plasma}
\label{sec4} 
In this section, we solve the dispersion relation derived in the previous section to calculate the damping rate of Alfv{\'e}n waves in the partially ionized plasma. 
In order to make physics more understandable, we take some limits. 
Since the proton cyclotron frequency is $\Omega_{\rm p}=10^{-2}~{\rm sec}^{-1} (B/1{\rm \mu G})$ and the collision frequencies are $\nu_{st} = 10^{-10}~{\rm sec}^{-1} ( n /1~{\rm cm}^{-3}) ( \sigma_{st} /10^{-16}~{\rm cm}^{-2})(v_{\rm rel}/10^6~{\rm cm~s}^{-1})$, the inequality of $\Omega_{\rm p} \gg \nu_{{\rm v},st}$ is satisfied in the typical interstellar medium in the current universe. 
In this case, from Eq.(\ref{eq:ltaup}), $\omega {\bm k} \cdot \delta {\bm \tau} \sim (k^2c_{\rm s}^2/\Omega_{\rm p})\rho_0\delta {\bm u}$ and from the first and second terms on the left hand side of Eq.(\ref{eq:lup}), the damping rate due to the proton viscosity is estimated as $\gamma_{\rm d}\sim k^2c_{\rm s}^2/\Omega_{\rm p}$. 
Since this viscous damping rate is much smaller than the frictional damping rate, we can ignore the effect of the proton viscosity, ${\bm k} \cdot \delta {\bm \tau}$ \citep{kulsrud69}. 
Thus, we do not need to solve Eq.(\ref{eq:ltaup}). 
Then, the dispersion relation is reduced to 
\begin{equation}
D=\omega^2 -k^2v_{\rm A}^2 - i\omega \nu_{\rm pH} + \frac{i\omega \nu_{\rm pH}^2\rho_0}{i\omega \rho_{\rm H,0} +\nu_{\rm pH}\rho_0+k^2c_{\rm s}^2\rho_{\rm H,0}/(i\omega+\nu_{\rm v,HH})}=0. 
\label{eq:rd1}
\end{equation}
The first and second terms originate from the inertia term of plasma fluid and the magnetic tension force, respectively. 
The last two terms originate from the friction term, that is, the right hand side of Eq. (\ref{eq:lup}). 
In this work for $\Omega_{\rm p} \gg \nu_{{\rm v},st}$, at first, we consider low and high frequency limits ($\omega \ll \nu_{\rm v,HH}$ and $\omega \gg \nu_{\rm v,HH}$). 
Then, we consider three cases, highly ionized plasma ($\rho_0/\rho_{\rm H,0} \gg1$), weakly ionized plasma ($\rho_0/\rho_{\rm H,0} \ll1$), and mildly ionized plasma ($\rho_0/\rho_{\rm H,0} \sim 1$).

On the other hand, the cyclotron frequency is expected to be smaller than collision frequencies ($\Omega_{\rm p} \ll \nu_{{\rm v},st}$ ) in the dense region in the current universe or in the early universe in which the magnetic field is expected to be very small compared with that in the current universe. Then, the second term on the left hand side of Eq.(\ref{eq:ltaup}) can be ignored. 
As a result, the dispersion relation is reduced to 
\begin{equation}
D=\omega^2 -k^2v_{\rm A}^2 - i\omega (\nu_{\rm pH} + k^2c_{\rm s}^2 T_{11})+ \frac{i\omega \rho_0(\nu_{\rm pH}-k^2c_{\rm s}^2T_{12})( \nu_{\rm pH}-k^2c_{\rm s}^2T_{21} )}{i\omega \rho_{\rm H,0}+(\nu_{\rm pH}+k^2c_{\rm s}^2T_{22})\rho_0}=0,~
\label{eq:rd2}
\end{equation}
where $T_{11}, T_{12},T_{21}$, and $T_{22},$ are given by
\begin{equation}
    \left(
    \begin{array}{cc}
    T_{11} & T_{12} \\
    T_{21} & T_{22} 
    \end{array}
    \right)
  = \frac{1}{(i\omega+\nu_{\rm v,pp})(i\omega+\nu_{\rm v,HH})-\nu_{\rm v,pH}\nu_{\rm v,Hp}}
      \left(
    \begin{array}{cc}
    i\omega+\nu_{\rm v,HH} & \nu_{\rm v,pH}(\rho_{\rm H,0}/\rho_0) \\
    \nu_{\rm v,Hp} & (i\omega+\nu_{\rm v,pp})(\rho_{\rm H,0}/\rho_0) 
    \end{array}
    \right)
\label{eq:tt}
\end{equation}
Since we focus on Alfv{\'e}n waves in this work, the frequency is smaller than the cyclotron frequency. 
Therefore, in this work for $\Omega_{\rm p} \ll \nu_{{\rm v},st}$, we consider only for the low frequency limit ($\omega \ll \Omega_{\rm p} \ll \nu_{{\rm v},st}$), in which case $\omega$ can be ignored in Eq.(\ref{eq:tt}). 

\subsection{For $\omega \ll \nu_{\rm v,HH}$ and $\Omega_{\rm p}/\nu_{{\rm v},st}\gg1$}
\label{sec:sub1}
In the low frequency limit ($\omega \ll \nu_{\rm v,HH}$), the damping rate and the real part of frequency are expected to be proportional to $k^2$ and $k$, so that the damping rate is expected to be smaller than the real part of frequency. 
Then, by using the weak damping approximation, the real and imaginary parts of the dispersion relation can be reduced to 
\begin{equation}
\rm{Re}[D(\omega_{\rm r},k)] = 0 ; ~~ \gamma_{\rm d} =-\frac{\rm{Im}[D(\omega_{\rm r},k)]}{\frac{\partial \rm{Re}[D(\omega_{\rm r},k)]}{\partial \omega_{\rm r}}},
\label{eq:wda}
\end{equation}
where $\omega_{\rm r}=\rm{Re}[\omega]$ is the real part of the frequency. 
Inserting Eq. (\ref{eq:rd1}) to Eq. (\ref{eq:wda}) and taking the low frequency limit, we obtain the real and imaginary parts of the frequency
\begin{eqnarray}
\omega_{\rm r,low} &=& \pm kv_{\rm A}  \sqrt{\frac{\rho_0}{\rho_0+\rho_{\rm H,0}} }~~,\label{eq:lowdis2} \\
\gamma_{\rm d,low} &=& \frac{k^2}{2}  \left \{ \frac{v_{\rm A}^2}{\nu_{\rm pH}} \left( \frac{\rho_{\rm H,0}}{\rho_0+\rho_{\rm H,0}} \right)^2+\frac{c_{\rm s}^2}{\nu_{\rm v,HH}} \left( \frac{\rho_{\rm H,0}}{\rho_0+\rho_{\rm H,0}} \right) \right\}~~, 
\label{eq:lowdis}
\end{eqnarray}
where $v_{\rm A}\sqrt{\rho_0/(\rho_0+\rho_{\rm H,0})}$ is the Alfv{\'e}n velocity defined by the total density. 
In the low frequency limit, the hydrogen gas and plasma gas frequently interact with each other, so that they almost behave as a single fluid. 
As shown in the black solid, dashed, and dotted lines in Figs.\ref{fig1}-\ref{fig3}, the above equation solutions are in good agreement with the numerical solutions to Eq.(\ref{eq:fd}) (cyan curves) in the low frequency limit. 
The first and second terms on the damping rate are due to the friction between plasma and hydrogen gas, and viscosity of hydrogen gas, respectively. 
The ratio of the viscous damping rate to the frictional damping rate is 
\begin{eqnarray}
\frac{\gamma_{\rm d,low,vis}}{\gamma_{\rm d,low,fri}} = \frac{5\beta_{\rm p}}{2}\frac{\rho_0+\rho_{\rm H,0}}{4\rho_0+3\rho_{\rm H,0}(\sigma_{\rm HH}/\sigma_{\rm pH})} \approx
\beta_{\rm p} \times \left\{ \begin{array}{ll}
5/8 & (~{\rm for}~\rho_0 \gg \rho_{\rm H,0}~) \\
 (5/6)(\sigma_{\rm pH}/\sigma_{\rm HH})& (~{\rm for}~\rho_0 \ll \rho_{\rm H,0}~)
 \end{array} \right. ~~.
\end{eqnarray}
where $\beta_{\rm p} = 8\pi p_{\rm p,0}/B_0^2$ is the proton beta defined by the proton pressure,  and $\sigma_{\rm pH}/\sigma_{\rm HH}$ is on the order of unity. 
As shown in the last equation, the viscosity of hydrogen gas significantly contributes the damping of Alfv{\'e}n wave as long as $\beta_{\rm p}$ is larger than on the order of unity. 
Note that the above ratio does not strongly depend on the ionization fraction directly. 

\subsection{For $\omega \gg \nu_{\rm v,HH}$ and $\Omega_{\rm p}/\nu_{{\rm v},st}\gg1$}
\label{sec:sub2}
In the high frequency limit ($\omega \gg \nu_{\rm v,HH}$), the damping rate and the real part of frequency are expected to be $0.5\nu_{\rm pH}$ and $kv_{\rm A}$, so that the damping rate is again expected to be smaller than the real part of frequency. 
Then, by using the weak damping approximation and taking the high frequency limit, we obtain the real and imaginary parts of the frequency
\begin{eqnarray}
\omega_{\rm r,high} &=& \pm kv_{\rm A}, \label{eq:highdis0}\\
\gamma_{\rm d,high} &=& \frac{\nu_{\rm pH}}{2} \times \left\{ \begin{array}{ll}
 \frac{\nu_{\rm v,HH}\rho_{\rm H,0}}{\nu_{\rm pH}\rho_0+\nu_{\rm v,HH}\rho_{\rm H,0}} & (~{\rm for}~ v_{\rm A} = c_{\rm s}~) \\
 1& (~{\rm for}~v_{\rm A} \neq c_{\rm s}~)
 \label{eq:highdis1}
\end{array} \right. \\
&=& \frac{\nu_{\rm pH}}{2} \times \left\{ \begin{array}{ll}
 \frac{8+6(\sigma_{\rm HH}/\sigma_{\rm pH})(\rho_{\rm H,0}/\rho_0)}{13+6(\sigma_{\rm HH}/\sigma_{\rm pH})(\rho_{\rm H,0}/\rho_0)} & (~{\rm for}~ v_{\rm A} = c_{\rm s}~) \\
 1& (~{\rm for}~v_{\rm A} \neq c_{\rm s}~)
\label{eq:highdis2}
\end{array} \right. ~~.
\end{eqnarray}
For $v_{\rm A} \neq c_{\rm s}$, the damping rate is given only by the friction, that is, the viscosity and pressure tensor of hydrogen gas do not affect on the damping of the Alfv{\'e}n wave. 
On the other hand, for the special case of $v_{\rm A} = c_{\rm s}$, the pressure tensor of hydrogen gas slightly changes the damping rate, $\gamma_{\rm d,high}\approx (4/13)\nu_{\rm pH}$ for $\rho_{\rm H,0}/\rho_0\ll 1$ and $\gamma_{\rm d,high}\approx (1/2)\nu_{\rm pH}$ for $\rho_{\rm H,0}/\rho_0\gg 1$. 
As shown in the black solid, dashed, and dotted lines in Figs.\ref{fig1}-\ref{fig3}, the above approximate solutions are in good agreement with the numerical solutions in the high frequency limit. 

\subsection{For $\rho_0/\rho_{\rm H,0} \gg1$ and $\Omega_{\rm p}/\nu_{{\rm v},st}\gg1$}
\begin{figure}
\plottwo{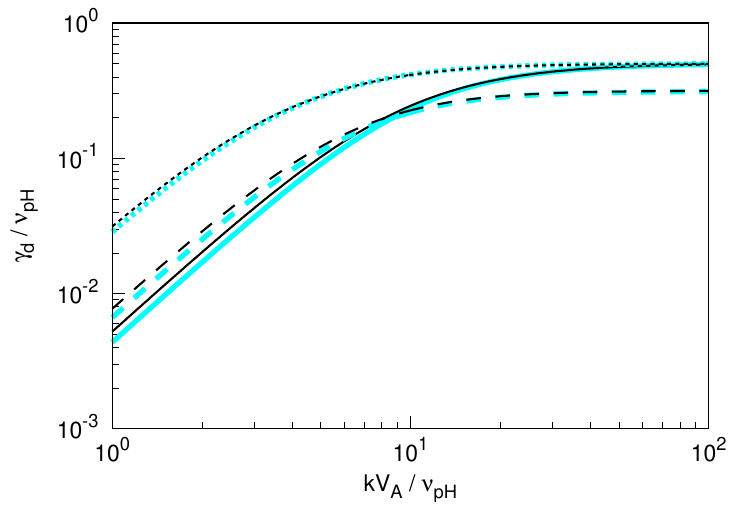}{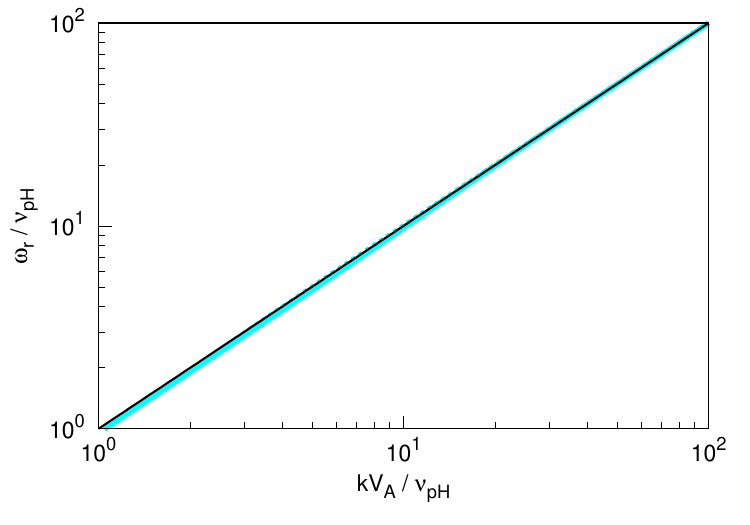}
\caption{Damping rate, $\gamma_{\rm d}=\rm{Im}[\omega]$ (left) and real part of frequency, $\omega_{\rm r}=\rm{Re}[\omega]$ (right) for $\rho_0/\rho_{\rm H,0}=10$ and $\Omega_{\rm p}/\nu_{\rm pH}=10^8$. The solid, dashed, and dotted curves show results for $\beta_{\rm p}=0.1,1, {\rm and}~10$, respectively. The cyan and black curves show numerical solutions to Eq.(\ref{eq:fd}) and approximate analytical solutions, Eqs. (\ref{eq:ana1}) and (\ref{eq:ana2}), respectively.}
\label{fig1}
\end{figure}

For a highly ionized plasma ($\rho_0/\rho_{\rm H,0} \gg1$), the friction force acting on the plasma fluid is much smaller than the tension force of the magnetic field, that is, the first two terms in Eq. (\ref{eq:rd1}) are much larger than the last two terms. 
Therefore, the damping rate should be smaller than the frequency of the Alfv{\'e}n wave, $\rm{Im}[\omega] \ll \rm{Re}[\omega]$. 
Then, the real and imaginary parts of the frequency are given by using the weak damping approximation 
%
\begin{eqnarray}
\omega_{\rm r}&=&\pm kv_{\rm A}~~, \label{eq:ana1}\\
\gamma_{\rm d} &=&\frac{\nu_{\rm pH}}{2}\left[1-\frac{  (k^2v_{\rm A}^2+\nu_{\rm v,HH}^2)\nu_{\rm pH}^2\rho_0^2 + k^2c_{\rm s}^2\nu_{\rm v,HH}\nu_{\rm pH}\rho_0\rho_{\rm H,0}}{  \{k^2(c_{\rm s}^2 -v_{\rm A}^2)\rho_{\rm H,0}  + \nu_{\rm v,HH}\nu_{\rm pH}\rho_0\}^2 + k^2v_{\rm A}^2(\rho_0\nu_{\rm pH}+\rho_{\rm H,0}\nu_{\rm v,HH})^2}\right]~~,
\label{eq:ana2}
\end{eqnarray}
where the friction terms in Eq. (\ref{eq:rd1}) is ignored for solving the real part of the dispersion relation, and the contributions from the friction term on the denominator, $\partial \rm{Re}[D(\omega_{\rm r},k)]/\partial \omega_{\rm r}$, is ignored for solving the imaginary part of the dispersion relation. 
In the low frequency limit, the above damping rate can be represented by a simpler form, 
\begin{equation}
\gamma_{\rm d,low} =\frac{k^2}{2} \left \{ \frac{ v_{\rm A}^2}{\nu_{\rm pH}} \left( \frac{\rho_{\rm H,0}}{\rho_0} \right)^2 + \frac{c_{\rm s}^2}{\nu_{\rm v,HH}} \left( \frac{\rho_{\rm H,0}}{\rho_0} \right) \right \}.
\end{equation}
This expression coincides with Eq. (\ref{eq:lowdis}) for $\rho_0/\rho_{\rm H,0} \gg1$.
In the high frequency limit, the damping rate can be also represented by a simpler form, 
\begin{eqnarray}
\gamma_{\rm d,high} &=& \frac{\nu_{\rm pH}}{2} \times \left\{ \begin{array}{ll}
 \frac{\nu_{\rm v,HH}\rho_{\rm H,0}}{\nu_{\rm pH}\rho_0+\nu_{\rm v,HH}\rho_{\rm H,0}} & (~{\rm for}~ v_{\rm A} = c_{\rm s}~) \\
 1& (~{\rm for}~v_{\rm A} \neq c_{\rm s}~)\\
\end{array} \right. ~~.
\label{eq:hw}
\end{eqnarray}
This expression again coincides with Eq. (\ref{eq:highdis1}).

Fig.~\ref{fig1} shows the damping rate (left panel) and real part of the frequency (right panel) for $\rho_0/\rho_{\rm H,0}=10$ and $\Omega_{\rm p}/\nu_{\rm pH}=10^8$, where the cyan and black curves are numerical solutions to Eq.(\ref{eq:fd}) and approximate analytical solutions, Eqs. (\ref{eq:ana1}) and (\ref{eq:ana2}), respectively. 
The solid, dashed, and dotted curves show results for $\beta_{\rm p}=0.1,1, {\rm and}~10$, respectively. 
As one can see, our approximate analytical solutions are in good agreement with the numerical solutions. 

\subsection{For $\rho_0/\rho_{\rm H,0} \ll1$ and $\Omega_{\rm p}/\nu_{{\rm v},st}\gg1$}
\begin{figure}
\plottwo{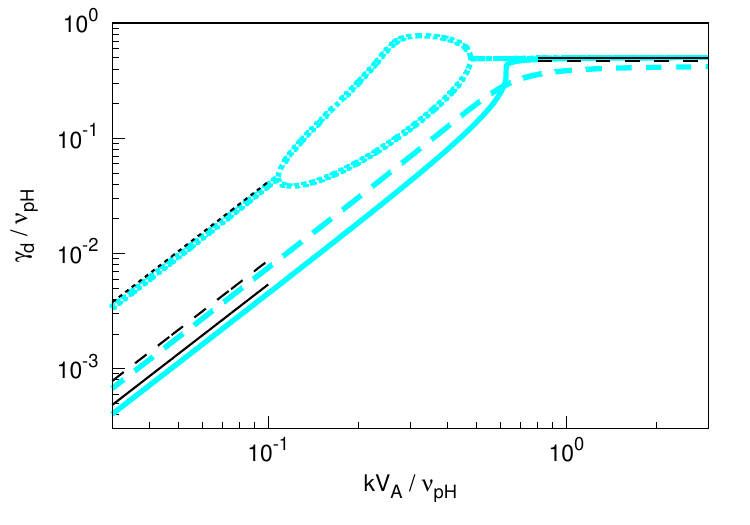}{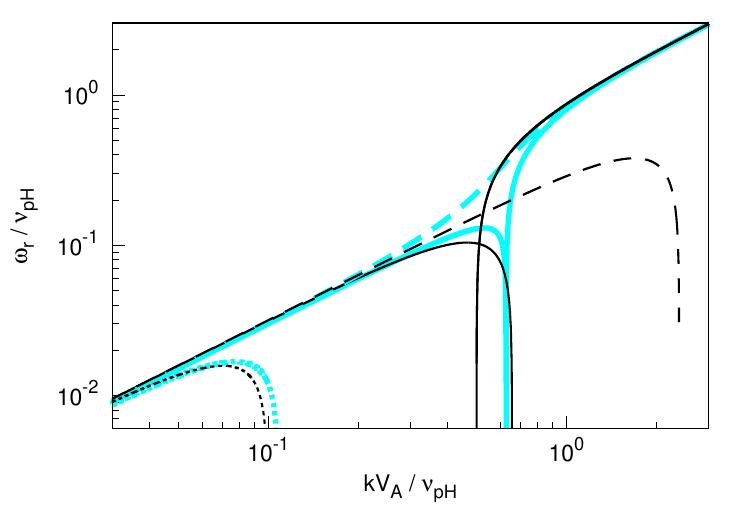}
\caption{The same as Fig.~\ref{fig1}, but for $\rho_0/\rho_{\rm H,0}=0.1$. The black curves show approximate analytical solutions, Eqs. (\ref{eq:rw1}) and (\ref{eq:rw12}) for the low frequency limit, and Eqs. (\ref{eq:rw2}) and (\ref{eq:highdis2}) for the high frequency limit.}
\label{fig2}
\end{figure}

For a weakly ionized plasma ($\rho_0/\rho_{\rm H,0} \ll1$), the weak damping approximation cannot be applied except for the high and low frequency limits because the friction force can be comparable to the magnetic tension force and the damping rate due to the hydrogen viscosity can be comparable to the frequency of the Alfv{\'e}n wave. 
In the low frequency range ($\omega \ll \nu_{\rm pH}$), the plasma gas tightly couples with the hydrogen gas. 
In this case, since the inertia of the plasma gas can be ignored for $\rho_0/\rho_{\rm H,0} \ll1$, 
the first term of Eq.(\ref{eq:rd1}) can be ignored. 
Then, the dispersion relation of Eq.(\ref{eq:rd1}) is reduced to a quadratic equation in $\omega$. 
The solution is 
\begin{eqnarray}
\omega_{\rm r,low} &=&  \pm \sqrt{k^2v_{\rm A}^2 \left( \frac{\rho_0}{\rho_{\rm H,0}} \right) -\frac{k^4}{4} \left( \frac{v_{\rm A}^2}{\nu_{\rm pH}} - \frac{c_{\rm s}^2}{\nu_{\rm v,HH}}  \right)^2 }, \label{eq:rw1}\\
\gamma_{\rm d,low} &=&\frac{k^2}{2} \left( \frac{ v_{\rm A}^2}{\nu_{\rm pH}}  + \frac{c_{\rm s}^2}{\nu_{\rm v,HH}} \right). 
\label{eq:rw12}
\end{eqnarray}
On the other hand, in the high frequency limit ($\omega \gg \nu_{\rm pH}$), the plasma gas rarely interacts with hydrogen gas during the oscillation period, so that the hydrogen does not move much, $\delta \bm{u}_{\rm H}\approx \bm{0}$.
Then, the last term of Eq.(\ref{eq:rd1}) can be ignored, so that the dispersion relation of Eq.(\ref{eq:rd1}) is reduced to a quadratic equation in $\omega$. 
The solution is 
\begin{eqnarray}
\omega_{\rm r,high} &=&  \pm \sqrt{k^2v_{\rm A}^2 -\frac{\nu_{\rm pH}^2}{4} }, \label{eq:rw2}\\
\gamma_{\rm d,high} &=&\frac{\nu_{\rm pH}}{2} ~~.  
\label{eq:rw22}
\end{eqnarray}
The above damping rates for low and high frequency limits coincide with Eqs. (\ref{eq:lowdis}) and (\ref{eq:highdis1}) for $\rho_0/\rho_{\rm H,0} \ll1$.

Eqs.(\ref{eq:rw1}) and (\ref{eq:rw2}) show that the real part of the frequency becomes zero, $\omega_{r}=0$, in the wavenumber region of $k_{\rm cr,low}\leq k\leq k_{\rm cr,high}$, where $k_{\rm cr,low}$ and $k_{\rm cr,high}$ are 
\begin{equation}
k_{\rm cr,low} =  \frac{2\nu_{\rm pH}}{v_{\rm A}}\frac{(\rho_0/\rho_{\rm H,0})^{1/2}}{\left| 1 - \left(\frac{\nu_{\rm pH}}{\nu_{\rm v,HH}}\right) \left( \frac{c_{\rm s}}{v_{\rm A}}\right)^2\right|} ;~~k_{\rm cr,high} = \frac{\nu_{\rm pH}}{2v_{\rm A}} ~~.  
\end{equation}
These critical wave numbers without viscosity coincide with results of \citet{soler13,soler24}. 
Alfv{\'e}n waves are strongly damped in this wavenumber region. 
In our approximated analysis, the strong damping region appears for $\rho_0/\rho_{\rm H,0}<1/16$ when the viscosity of hydrogen atom is ignored, which is the same as the result in \citet{kulsrud69}. 
The more accurate analysis without viscosity shows that this critical proton fraction is $1/8$ \citep{zweibel82,soler13,soler24}. 
For a sufficiently large proton beta ($\beta_{\rm p}=c_{\rm s}^2/v_{\rm A}^2 \gg \nu_{\rm v,HH}/\nu_{\rm pH}$), the viscosity of hydrogen atoms makes the strong damping region and the critical proton fraction larger. 
On the other hand, for the case of $\beta_{\rm p} \sim \nu_{\rm v,HH}/\nu_{\rm pH}$, the critical wavenumber in the low frequency side, $k_{\rm cr, low}$, becomes larger than that in the high frequency side, $k_{\rm cr,high}$, resulting in the absence of the strong damping region ($\omega_{\rm r}=0$).

Fig.~\ref{fig2} shows the damping rate (left panel) and real part of the frequency (right panel) for $\rho_0/\rho_{\rm H,0}=0.1$ and $\Omega_{\rm p}/\nu_{\rm pH}=10^8$, where the cyan and black curves are numerical solutions to Eq.(\ref{eq:fd}) and approximate analytical solutions, Eqs. (\ref{eq:rw1}) and (\ref{eq:rw12}) for the low frequency limit, and Eqs. (\ref{eq:rw2}) and (\ref{eq:rw22}) for the high frequency limit. 
The solid, dashed, and dotted curves show results for $\beta_{\rm p}=0.1,1, {\rm and}~10$, respectively. 
As one can see, our approximate analytical solutions are in good agreement with the numerical solutions in low and high frequency limits. 
\subsection{For $\rho_0/\rho_{\rm H,0} \sim1$ and $\Omega_{\rm p}/\nu_{{\rm v},st}\gg1$}
\begin{figure}
\plottwo{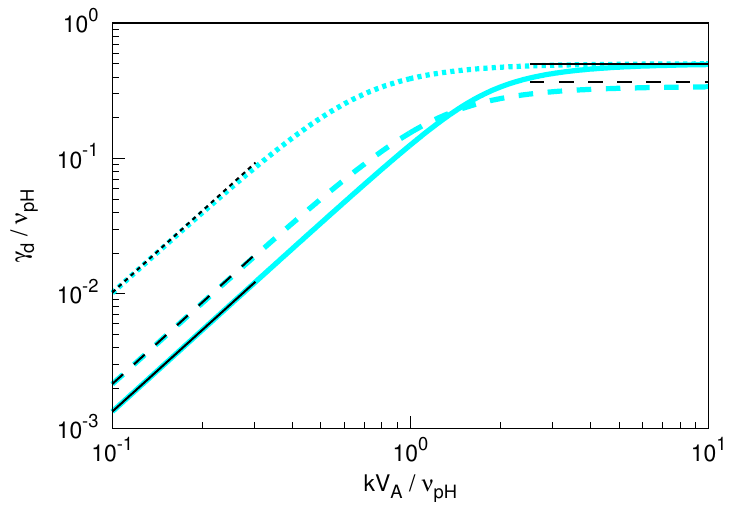}{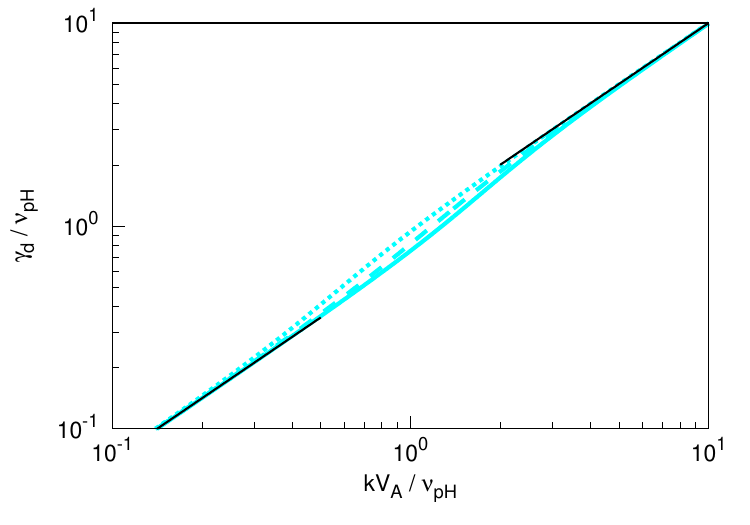}
\caption{The same as Fig.~\ref{fig1}, but for $\rho_0/\rho_{\rm H,0}=1$. The black curves show approximate analytical solutions, Eqs. (\ref{eq:lowdis2}) and (\ref{eq:lowdis}) for the low frequency limit, and Eqs. (\ref{eq:highdis0}) and (\ref{eq:highdis2}) for the high frequency limit.}
\label{fig3}
\end{figure}

For a mildly ionized plasma ($\rho_0/\rho_{\rm H,0} \sim1$), we cannot apply any approximations except for the high and low frequency limits. 
Fig.~\ref{fig3} shows the damping rate (left panel) and real part of the frequency (right panel) for $\rho_0/\rho_{\rm H,0}=1$ and $\Omega_{\rm p}/\nu_{\rm pH}=10^8$, where the cyan and black curves are numerical solutions to Eq.(\ref{eq:fd}) and 
approximate analytical solutions, Eqs. (\ref{eq:lowdis2}) and (\ref{eq:lowdis}) for the low frequency limit, and Eqs. (\ref{eq:highdis0}) and (\ref{eq:highdis2}) for the high frequency limit. 
The solid, dashed, and dotted curves show results for $\beta_{\rm p}=0.1,1, {\rm and}~10$, respectively. 
As one can see, our approximated analytical solutions are in good agreement with the numerical solutions in the high and low frequency limits. 

\subsection{For $\omega \ll \nu_{\rm v,HH}$ and $\Omega_{\rm p}/\nu_{{\rm v},st}\ll1$}
As discussed in section \ref{sec:sub1}, the weak damping approximation can be applied in the low frequency limit even for $\Omega_{\rm p}/\nu_{{\rm v},st}\ll1$. 
Inserting Eq. (\ref{eq:rd2}) to Eq. (\ref{eq:wda}) and taking the low frequency limit, we obtain the real and imaginary parts of the frequency, 
\begin{eqnarray}
\omega_{\rm r,low} &=& \pm kv_{\rm A}  \sqrt{\frac{\rho_0}{\rho_0+\rho_{\rm H,0}} } \label{eq:lowdis21}~~,\\
\gamma_{\rm d,low} &=& \frac{k^2}{2}  \left \{ \frac{v_{\rm A}^2}{\nu_{\rm pH}} \left( \frac{\rho_{\rm H,0}}{\rho_0+\rho_{\rm H,0}} \right)^2+ \frac{c_{\rm s}^2}{\nu_{\rm v,tot}}\left( \frac{\rho_{\rm H,0}}{\rho_0+\rho_{\rm H,0}} \right) \right\}~~, 
\label{eq:lowdis22}
\end{eqnarray}
where $\nu_{\rm v,tot}=(\rho_{\rm H,0}/\rho_0)/(T_{11}+T_{12}+T_{21}+T_{22})$ is 
\begin{equation}
\nu_{\rm v,tot} =\frac{2}{5}\frac{(3\nu_{\rm pp}+4\nu_{\rm pH})(3\nu_{\rm HH}+4\nu_{\rm Hp})-\nu_{\rm Hp}\nu_{\rm pH}}{(3\nu_{\rm HH}+4\nu_{\rm Hp})(\rho_0/\rho_{\rm H,0})+3\nu_{\rm pp}+4\nu_{\rm pH}+2\nu_{\rm Hp}}~~.
\end{equation}
$\nu_{\rm pp}=10^{-5}~{\rm sec} (n_{\rm p}/1{\rm cm}^{-3})(T/1{\rm eV})^{-3/2}$ is the collision frequency due to the Coulomb interaction, which is usually much larger  than other collision frequencies due to atomic processes, $\nu_{st} = 10^{-10}~{\rm sec}^{-1} ( n /1~{\rm cm}^{-3}) ( \sigma_{st} /10^{-16}~{\rm cm}^{-2})(v_{\rm rel}/10^6~{\rm cm~s}^{-1})$. 
In this case, this effective collision frequency, $\nu_{\rm v,tot}$, is almost the same as $\nu_{\rm v,HH}$, so that the damping rate given in Eq. (\ref{eq:lowdis22}) is also almost the same as the damping rate for $\Omega_{\rm p}/\nu_{{\rm v},st}\gg1$, Eq.(\ref{eq:lowdis}). 

For the low frequency limit in a weakly ionized plasma, the inertia of the plasma component can be ignored, so that the first term of Eq.(\ref{eq:rd2}) can be ignored. 
Then, the dispersion relation, Eq.(\ref{eq:rd2}), becomes a quadratic equation in $\omega$. 
The solution is given by
\begin{eqnarray}
\omega_{\rm r,low} &=&  \pm \sqrt{k^2v_{\rm A}^2 \left( \frac{\rho_0}{\rho_{\rm H,0}} \right)\left\{1+\frac{k^2c_{\rm s}^2}{\nu_{\rm pH}}(T_{11}+T_{22})\right\} -\frac{k^4}{4} \left( \frac{v_{\rm A}^2}{\nu_{\rm pH}} + \frac{c_{\rm s}^2}{\nu_{\rm v,tot}}  \right)^2 }, \label{eq:rw3} \label{eq:ddd}\\
\gamma_{\rm d,low} &=&\frac{k^2}{2}  \left ( \frac{v_{\rm A}^2}{\nu_{\rm pH}} + \frac{c_{\rm s}^2}{\nu_{\rm v,tot}} \right).  
\end{eqnarray}
As mentioned above, $\nu_{\rm pp}$ is usually much larger  than other collision frequencies. 
In this case, $\nu_{\rm v,tot}\approx\nu_{\rm v,HH}$ and $T_{11}+T_{22}\approx (\rho_{\rm H,0}/\rho_0)/\nu_{\rm v,HH}$, so that the real part of the frequency, Eq. (\ref{eq:ddd}), is almost the same as that for $\Omega_{\rm p}/\nu_{{\rm v},st}\gg1$, Eq.(\ref{eq:rw1}). 
The above damping rate and real part of the frequency in low frequency limit coincide with Eqs. (\ref{eq:lowdis21}) and (\ref{eq:lowdis22}) for $\rho_0/\rho_{\rm H,0}\ll1$.
Eq.(\ref{eq:rw3}) shows that the real part of the frequency becomes zero, $\omega_{r}=0$, for $k\geq k_{\rm cr,low}$, where $k_{\rm cr,low}$ is 
\begin{equation}
k_{\rm cr,low} =  \frac{2\nu_{\rm pH}}{v_{\rm A}}\frac{(\rho_0/\rho_{\rm H,0})^{1/2}}{\left| \left\{1+ \left(\frac{\nu_{\rm pH}}{\nu_{\rm v,tot}}\right) \left( \frac{c_{\rm s}}{v_{\rm A}}\right)^2\right\}^2-\left( \frac{c_{\rm s}}{v_{\rm A}}\right)^2 \left( \frac{\rho_0}{\rho_{\rm H,0}}\right)\nu_{\rm pH}(T_{11}+T_{22})  \right|^{1/2}} ~~.  
\end{equation}
This critical wave number without viscosity coincides with results of \citet{soler13,soler24}. 
Alfv{\'e}n waves are heavy damped in the wavenumber range of $k\geq k_{\rm cr,low}$. 
Note that the wavenumber should be much smaller than $\Omega_{\rm p}/v_{\rm A}$ because we focus on Alfv{\'e}n waves in this work. 

Although we do not show the numerical solution, we confirmed that the above approximate analytical solutions for $\Omega_{\rm p}/\nu_{{\rm v},st}\ll1$ are in good agreement with the numerical solution to Eq.(\ref{eq:fd}).

\section{Discussion}
\label{sec5}
In the section~\ref{sec:sub2}, we showed that the viscosity does not significantly change the damping rate of the Alfv{\'e}n wave for the high frequency limit, $\omega \gg \nu_{\rm v,HH}$. 
In the opposite limit, $\omega \ll \nu_{\rm v,HH}$, our analysis in the section~\ref{sec:sub1} showed the viscosity enhances the damping rate of the Alfv{\'e}n wave in the high beta plasma. 
Furthermore, in a weakly ionized plasma ($\rho_0/\rho_{\rm H,0}\ll1$), the viscosity makes the strong damping region ($\omega_{\rm r}=0$) large in the wavenumber space, which affects the energy spectrum of particles accelerated by a shock in a weakly ionized plasma \citep{malkov11}.
In addition to collisions between hydrogen atoms, collisions between hydrogen atoms and protons also affect the viscosity of hydrogen atoms. 
Therefore, the viscosity of hydrogen atoms could contribute the damping of the Alfv{\'e}n wave even in a highly ionized plasma. 
The plasma $\beta$ is expected to be large in the shock downstream region, current sheet, intergalactic medium, plasma in the early universe, and plasma in the laboratory. 
In these plasmas, the Alfv{\'e}n wave could mainly dissipate due to the viscosity of hydrogen atoms.  
  
In addition to the viscous and frictional dampings, the Alfv{\'e}n wave decays owing to turbulence \citep{farmer04,yan04}, three-wave interaction \citep{sagdeev69}, and the resonance between thermal ions and beat waves of Alfv{\'e}n waves \citep{lee73}. 
Since the damping rates of these nonlinear processes depend on the strength and spectrum of the magnetohydrodynamic turbulence, the dominant damping process depends on a specific situation. 
Qualitatively, in a partially ionized system with a high plasma $\beta$ and weak turbulence, the Alfv{\'e}n wave mainly decays due to the viscous damping.

It should be noted that even in a low $\beta$ plasma, Alfv{\'e}n waves propagating obliquely to a magnetic field line are dissipated by the hydrogen viscosity \citep{lazarian04} because the frequency, $\omega = kv_{\rm A} \cos \theta$, becomes smaller, where $\theta$ is the angle between the magnetic field line and the wavevector.

To derive the damping rate of the Alfv{\'e}n wave propagating along a magnetic field line, we have applied the Grad's ten-moment model in this work (\ref{sec2}). 
To calculate the damping rates of fast and slow waves, the ten-moment model has to be extended to thirteen-moment model that incorporates the heat conduction. 
Furthermore, for waves propagating obliquely to a magnetic field line, the traceless part of the pressure tensor and heat conduction are expressed in very complicated formulae. 
In addition to hydrogen atoms, helium atoms also could play an important role in some cases \citep{recchia22}. 
A comprehensive analysis will be addressed in future work. 

In the Grad's method, the deviation from the Maxwell distribution is expanded in tensorial Hermit polynomial (Eq.(\ref{eq:hermite})). 
However, this is not always a sufficiently accurate approximation. 
In particular, kinetic simulations show that the momentum distribution is more complicated around a collisionless shock in the partially ionized plasma \citep{ohira13b,ohira16a,ohira16b}. 
Although it is challenging, in order to understand the acceleration and propagation of CRs in the partially ionized plasma, the decay of MHD waves in such a system should also be addressed. 

\section{Summary}
\label{sec6}
In this work, we have revisited the damping of Alfv{\'e}n waves in partially ionized plasmas. 
In the CR community, the friction force between ions and neutral particles has been considered as the main damping process of the Alfv{\'e}n wave, whereas the viscosity of hydrogen atoms has been considered in studies of interstellar MHD turbulence. 
It was not clear in what plasma Alfv{\'e}n waves are mainly dissipated by the viscosity.
Prior to this study, the viscosity was incorporated into the equation of motion for hydrogen gas in a phenomenological manner. 
In other words, the traceless part of the pressure tensor was implicitly assumed to be a quasi-steady state, that is, the time evolution of the traceless part of pressure tensor  was not solved. 
Although this is a good approximation in the low frequency limit ($\omega\ll \nu_{\rm st}$), the time derivative of the traceless part of the pressure tensor should be taken into account when the frequency is comparable to or larger than the scattering frequency. 
In this work, we have performed the linear analysis for the Alfv{\'e}n wave in the framework of the two-fluid ten-moment-equation system, where we have taken account of the time evolution of the traceless part of the pressure tensor. 
Furthermore, in addition to the interaction between hydrogen atoms, we have considered the interaction between protons and hydrogen atoms in the equation of the traceless part of the pressure tensor, which is the main cause of the viscosity  when the proton density is comparable to or larger than that of hydrogen atoms. 

We have numerically solved the dispersion relation for the Alfv{\'e}n wave propagating along a magnetic field line. 
In addition, by taking some limits, we have obtained some approximate analytical solutions. 
These solutions in the low frequency limit show that the viscosity enhances the damping rate of the Alfv{\'e}n wave in the high beta plasma (Eqs.(\ref{eq:lowdis}) and (\ref{eq:lowdis22})). 
On the other hand, the viscosity does not significantly affect the damping rate in the high frequency limit (\ref{eq:highdis2}). 
In a weakly ionized plasma ($\rho_0/\rho_{\rm H,0}\ll1$), the viscosity increases the strong damping region ($\omega_{\rm r}=0$) in a wavenumber space around $k\sim v_{\rm A}/\nu_{\rm pH}$. 

\begin{acknowledgments} 
We would like to thank Shu-ichiro Inutsuka for useful discussions. 
This work is supported by JSPS KAKENHI Grant Number JP25K00999 and JP26H00825. 
\end{acknowledgments}


\begin{thebibliography}{}
%
\bibitem[Aharonian \& Atoyan(1996)]{aharonian96}
Aharonian, F. A., \& Atoyan, A., 1996, \aap, 309, 917
%
\bibitem[Axford et al.(1977)]{axford77}
Axford, W. I., Leer, E., \& Skadron, G., 1977, Proc. 15th Int. Cosmic Ray Conf., Plovdiv, 11, 132
%
\bibitem[Bell(1978)]{bell78}
Bell, A. R., 1978, \mnras, 182, 147
%
\bibitem[Berezhko \& Krymskii(1981)]{berezhko81} 
Berezhko, E. G., \& Krymskii, G. F., 1981, Soviet Astron. Lett., 7, 352
%
\bibitem[Blandford \& Ostriker(1978)]{blandford78}
Blandford, R. D., \& Ostriker, J. P., 1978, \apj, 221, L29
%
\bibitem[Braginskii(1958)]{braginskii58}
Braginskii, S. I., 1958, JETP, 6, 358
%
\bibitem[Braginskii(1965)]{braginskii65}
Braginskii, S. I., 1965, RvPP, 1, 205
%
\bibitem[Brahimi et al.(2020)]{brahimi20}
Brahimi, L, Marcowith, A., \& Ptuskin, V. S., 2020, \aap, 633, A72
%
\bibitem[Burgers (1969)]{burgers69} 
Burgers, J. M., 1969, Flow Equations for Composite Gases (New York: Academic)
%
\bibitem[Chapman \& Cowling(1953)]{chapman53}
Chapman, S., \& Cowling, T. G., 1953, The Mathematical Theory of Nonuniform Gasses (Cambridge: Cambridge Univ. Press)
%
\bibitem[Earl et al.(1988)]{earl88} 
Earl, J. A., Jokipii, J. R. \& Morfill, G., 1988, \apj, 331, L91
%
\bibitem[Farmer \& Goldreich(2004)]{farmer04} 
Farmer, A. J., \& Goldreich, P., 2004, \apj, 604, 671
%
\bibitem[Fujita et al.(2009)]{fujita09} 
Fujita, Y., Ohira, Y., Tanaka, S. J., Takahara, F., 2009, \apj, 369, 1629
%
\bibitem[Gabici et al.(2009)]{gabici09} 
Gabici S., Aharonian F. A., Casanova S., 2009, \mnras, 369, 1629
%
\bibitem[Grad(1949a)]{grad49a}
Grad, H. 1949a, CPAM, 2, 325
%
\bibitem[Grad(1949b)]{grad49b}
Grad, H. 1949b, CPAM, 2, 331
%
\bibitem[Grad(1958)]{grad58}
Grad, H. 1958, HDP, 3, 205
%
\bibitem[Hunana et al. (2022)]{hunana22}
Hunana, P., et al., 2022, \apjs, 260, 26
%
\bibitem[Krymsky (1977)]{krymsky77}
Krymsky, G. F., 1977, Dokl. Akad. Nauk SSSR, 234, 1306
%
\bibitem[Kulsrud \& Pearce(1969)]{kulsrud69}
Kulsrud, R., \& Pearce, W. P., 1969, \apj, 156, 445
%
\bibitem[Lazarian et al.(2004)]{lazarian04} 
Lazarian, A., Vishniac, E. T., \& Cho, J., 2004, \apj, 603, 180
%
\bibitem[Lee \& V\"{o}lk (1973)]{lee73}
Lee, M. A., \& V\"{o}lk, H. J., 1973, Astrophys. Space Sci. 24, 31
%
\bibitem[Lemoine(2021)]{lemoine21}
Lemoine, M., 2021, \prd, 104, 063020
%
\bibitem[Li \& Chen(2010)]{li10}
Li, H., \& Chen, Y., 2010, \mnras, 409, L35
%
\bibitem[Malkov et al.(2011)]{malkov11}
Malkov, M. A., Diamond, P. H., \& Sagdeev, R. Z., 2011, Nature Communications, 2, 194
%
\bibitem[Ohira et al.(2010)]{ohira10}
Ohira, Y., Murase, K., \& Yamazaki, R., 2010, \aap, 513, A17 
%
\bibitem[Ohira et al.(2011)]{ohira11}
Ohira, Y., Murase, K., \& Yamazaki, R., 2011, \mnras, 410, 1577 
%
\bibitem[Ohira(2013a)]{ohira13a}
Ohira, Y., 2013a, \apjl, 767, L16
%
\bibitem[Ohira(2013b)]{ohira13b}
Ohira, Y., 2013b, \prl, 111, 245002 
%
\bibitem[Ohira(2016a)]{ohira16a}
Ohira, Y., 2016a, \apj, 817, 137 
%
\bibitem[Ohira(2016b)]{ohira16b}
Ohira, Y., 2016b, \apj, 827, 36
%
\bibitem[Ohira(2025)]{ohira25}
Ohira, Y., 2025, \mnras, 541, 2434 
%
\bibitem[Ostrowski(1990)]{ostrowski90} 
Ostrowski, M., 1990, \aap, 238, 435
%
\bibitem[Piddington(1956)]{piddington56} 
Piddington, J. H. 1956, \mnras, 116, 314
%
\bibitem[Ptuskin \& Zirakashvili(2005)]{ptuskin05} 
Ptuskin, V. S., \& irakashvili, V. N., 2005, \aap, 429, 755

\bibitem[Recchia et al.(2022)]{recchia22} 
Recchia, S., et al., 2022, \aap, 660, A57
%
\bibitem[Rieger \& Duffy(2006)]{rieger06} 
Rieger, F. M., \& Duffy, P., 2006, \apj, 652, 1044
%
\bibitem[Sagdeev \& Galeev(1969)]{sagdeev69} 
Sagdeev, R. Z., \& Galeev, A. A., 1969, Nonlinear Plasma Theory (New York: Benjamin)
%
\bibitem[Schlickeiser \& Miller(1998)]{schlickeiser98} 
Schlickeiser, R., Miller, J. A., 1998, \apj, 492, 352
%
\bibitem[Schunk(1975)]{schunk75} 
Schunk, R. W., 1975, P\&SS, 23, 437
%
\bibitem[Skilling(1975)]{skilling75} 
Skilling, J., 1975, \mnras, 172, 557
%
\bibitem[Soler et al.(2013)]{soler13} 
Soler, R., Carbonell, M., Ballester, J. L., \& Terradas, J., 2013, \apj, 767, 171
%
\bibitem[Soler et al.(2015)]{soler15} 
Soler, R., Carbonell, M., \& Ballester, J. L., 2015, \apj, 810, 146
%
\bibitem[Soler et al.(2024)]{soler24} 
Soler, R., 2024, Phil. Trans. R. Soc. A, 382, 20230223
%
\bibitem[Takamoto \& Inutsuka(2010)]{takamoto10} 
Takamoto, M., \& Inutsuka, S., 2010,  Progress of Theoretical Physics, 123, 903
%
\bibitem[Vranjes \& Krastic(2013)]{vranjes13} 
Vranjes, J., \& Krastic, P. S., 2013, \aap, 554, A22
%
\bibitem[Webb(1989)]{webb89} 
Webb, G. M., 1989, \apj, 340, 1112
%
\bibitem[Xu et al.(2015)]{xu15} 
Xu, S., Lazarian, A., \& Yan, H., 2015, \apj, 810, 44
%
\bibitem[Xu et al.(2016)]{xu16}  
Xu, S., Yan, H., \& Lazarian, A., 2016, \apj, 826, 166
%
\bibitem[Yan \& Lazarian (2004)]{yan04}  
Yan, H., \& Lazarian, A., 2004, \apj, 614, 757
%
\bibitem[Zweibel \& Shull.(1982)]{zweibel82}  
Zweibel, E. G., \& Shull, J. M., 1982, \apj, 826, 166
%
\end{thebibliography}
\end{document}